\documentclass[preprint2]{emulateapj}
\usepackage{multirow}

\newcommand{\kms}{km~s$^{-1}$}

\shorttitle{Stellar Dynamics and Metallicity of NGC~4449 and the stream}
\shortauthors{Toloba et al.}

\begin{document}

\title{New Surface Brightness Fluctuations Spectroscopic Technique: NGC~4449 and its Stellar Tidal Stream}


\author{Elisa~Toloba\altaffilmark{1,2}}\email{toloba@ucolick.org}
\author{Puragra~Guhathakurta\altaffilmark{1}}
\author{Aaron~J.~Romanowsky\altaffilmark{1,3}}
\author{Jean~P.~Brodie\altaffilmark{1}}
\author{David~Mart\'inez-Delgado\altaffilmark{4,5}}
\author{Jacob~A.~Arnold\altaffilmark{1}}
\author{Neel~Ramachandran\altaffilmark{6}}
\author{Kuriakose~Theakanath\altaffilmark{7}}
\affil{$^1$UCO/Lick Observatory, University of California, Santa Cruz, 1156 High Street, Santa Cruz, CA 95064, USA}
\affil{$^2$Texas Tech University, Physics Department, Box 41051, Lubbock, TX 79409-1051, USA}
\affil{$^3$Department of Physics \& Astronomy, San Jos\'e State University, One Washington Square, San Jose, CA 95192, USA}
\affil{$^4$Astronomisches Rechen-Institut, Zentrum f$\ddot{{\rm u}}$r Astronomie der Universit$\ddot{{\rm a}}$t Heidelberg, M$\ddot{{\rm o}}$nchhofstr 12-14, 69120, Heidelberg, Germany}
\affil{$^5$Max Plank Institut f$\ddot{{\rm u}}$r Astronomie, K$\ddot{{\rm o}}$nigsstuhl 17, 60117, Heidelberg, Germany}
\affil{$^6$Saint Francis High School, 1885 Miramonte Avenue, Mountain View, CA 94040, USA}
\affil{$^7$Bellarmine College Preparatory, 960~W Hedding Street, San Jose, CA 95126, USA}

\begin{abstract}

We present a new spectroscopic technique based in part on targeting the upward fluctuations of the surface brightness for studying the internal stellar kinematics and metallicities of low surface brightness galaxies and streams beyond the Local Group. The distance to these systems makes them unsuitable for targeting individual red giant branch (RGB) stars (tip of RGB at $I\gtrsim24$~mag) and their surface brightness is too low ($\mu_r\gtrsim 25$~mag~arcsec$^{-2}$) for integrated light spectroscopic measurements. This technique overcomes these two problems by targeting individual objects that are brighter than the tip of the RGB. We apply this technique to the star-forming dwarf galaxy NGC~4449 and its stellar stream. We use Keck/DEIMOS data to measure the line-of-sight radial velocity out to $\sim7$~kpc in the East side of the galaxy and $\sim8$~kpc along the stream. We find that the two systems are likely gravitationally bound to each other and have heliocentric radial velocities of $227.3\pm10.7$~\kms\ and $225.8\pm16.0$~\kms, respectively. Neither the stream nor the near half of the galaxy shows a significant velocity gradient. We estimate the stellar metallicity of the stream based on the equivalent width of its Calcium triplet lines and find [Fe/H]~$=-1.37\pm0.41$, which is consistent with the metallicity-luminosity relation for Local Group dwarf galaxies. Whether the stream's progenitor was moderately or severely stripped cannot be constrained with this metallicity uncertainty. We demonstrate that this new technique can be used to measure the kinematics and (possibly) the metallicity of the numerous faint satellites and stellar streams in the halos of nearby ($\sim 4$~Mpc) galaxies. 

\end{abstract}


\section{Introduction}

The $\Lambda$ cold dark matter ($\Lambda$CDM) cosmological scenario implies that galaxies assemble hierarchically, i.e. the smallest halos contribute to the build up of the most massive ones \citep[e.g.,][]{Springel06}. These smaller halos are severely affected by the larger potential well of the host and are tidally stripped or fully disrupted, appearing on the sky as substructures in form of streams and satellites \citep[e.g.,][]{Bullock05,Springel06,Cooper10}. Observationally, the detection of substructures in the halos of massive galaxies agrees with this scenario \citep[e.g.,][]{MD10,Atkinson13,Martin14}. However, there are still discrepancies between observations and simulations that challenge this galaxy formation model. For example, the number of observed satellites around massive galaxies is at least one order of magnitude smaller than the predicted number by cosmological simulations \citep[the so-called ``missing satellite problem'', e.g.,][]{Klypin99,Moore99}, and the number of dwarf satellites with high circular velocities is significantly smaller than the $\Lambda$CDM  predictions \citep[the so-called ``too big to fail problem'', e.g.,][]{BoylanKolchin11,BoylanKolchin12}.

These discrepancies between $\Lambda$CDM and observations are based almost entirely on studies of the satellites and streams found in the Local Group. However, models also predict a large dispersion in the number and properties of the satellites and streams due to varying accretion histories \citep{Johnston08} and inhomogeneous reionization \citep{Busha10}. Thus, it is important to make a systematic study of these faint structures beyond the Local Group and understand whether these inconsistencies between $\Lambda$CDM and observations are a consequence of the Local Group being an outlier or whether some of the physics considered in these simulations need to be adjusted. Some efforts have begun in this direction \citep{Chiboucas09,Chiboucas13,Sand14,Crnojevic14,etj16}, but they are generally limited to photometric information. Spectroscopy of these faint structures and satellites provides dynamical information to model the disruptive events and predict their time scales as well as stellar metallicity information to learn about their star formation histories. Spectroscopy can be carried out using globular clusters and planetary nebulae as bright dynamical tracers \citep[e.g.,][]{Foster14}. However, this approach is feasible only for the brightest dwarfs and streams because the faintest ones have very low numbers of these tracers.

Stellar radial velocity gradients and detailed stellar populations of galaxies have been analyzed using two main techniques: resolved stars and integrated light. The choice between these techniques is based on the distance of the target galaxy and its surface brightness. Low surface brightness galaxies within the Local Group, like the satellites of the Milky Way and M31, are studied using resolved stars because their proximity and the brightness of their individual stars allow them to be targeted spectroscopically with the current telescopes and reasonable integration times \citep[e.g.,][]{Geha06,SimonGeha07,Kirby08a,Kirby08b,Geha10,Kirby11,Tollerud12,Collins13,Ho15,Simon15}. Individual stars in galaxies outside the Local Group are too faint ($I\gtrsim 24$~mag) to be targeted spectroscopically with current telescopes. In those cases, the technique of integrated light spectroscopy is used. However, this can only be applied to relatively bright objects ($\mu_r <24$~mag~arcsec$^{-2}$), which limits the study of spatially resolved galaxy internal properties to nearby objects. Some examples of integrated light spectroscopic studies of spatially resolved dwarf galaxies are reported in \citet{Geha02,Geha03,DR05,Koleva09,etj09,etj11,etj14b,etj14a,etj15,Rys13}.
Here we describe a new spectroscopic technique that allows for kinematic and metallicity studies of faint satellites and streams beyond the Local Group and up to a distance of $\sim 4$~Mpc. 

\section{New SBF Spectroscopic Technique}\label{technique}

We have developed a new technique to spectroscopically target low surface brightness satellites and streams to obtain their internal stellar kinematics and (possibly) their metallicity. This technique combines two traditional spectroscopic methods: integrated light spectroscopy and multi-object spectroscopy of individual stars. The main idea is as follows: instead of placing a long slit across a low surface brightness target (e.g., dwarf galaxy or stellar stream), we break the long slit into multiple short segments (slitlets). These slitlets are then moved along the direction perpendicular to the length of the original long slit so they target the most prominent peaks of the non-uniform surface brightness distribution of the target galaxy or stream.

\subsection{Detailed description of the technique}\label{description}

This spectroscopic technique is based on the same phenomenon that is the basis of the surface brightness fluctuation (SBF) technique used to measure distances. Surface brightness fluctuations are caused by Poisson fluctuations in the number of stars blended within a seeing disk from one location to another. These fluctuations are the largest for the most luminous stars since their numbers are the smallest and yet their fractional contribution to the overall surface brightness is relatively high. The fractional amplitude of SBFs depends on distance, surface brightness, and stellar luminosity function in the following ways: (1) the further away a galaxy, the larger the number of stars in any given seeing disk and the smaller the Poission fluctuations or SBF; (2) the higher the surface brightness of the galaxy, the larger the number of stars in any seeing disk and the smaller the SBF (however, for high surface brightness objects the traditional long slit spectroscopy works better); and (3) the more steeply the star counts rise towards fainter magnitudes at the bright end of the luminosity function, the larger the SBF. The metallicity distribution function, the age distribution of the galaxy and the choice of the photometric filter can affect how steep the bright end of the luminosity function is.

We combine this new spectroscopy technique with the traditional spectral stacking resulting in two flavors of coaddition. First, since our technique targets upward SBFs, it essentially takes advantage of the natural blends of stars. Stars that are nearly co-spatial in projection are detected as bright blended sources in photometric catalogs derived from images. Of course, the apparent brightness of a blended source is higher than the brightnesses of the constituent individual stars. Second, we do the usual spectral coaddition of faint sources in order to boost the S/N ratio of the resulting spectrum. Next we describe the spectroscopic target selection and analyze these two different kinds of coaddition in the context of the application of our technique to the recently discovered stellar stream near the star-forming dwarf galaxy NGC~4449.

\subsection{Photometric selection of spectroscopic candidates}

Our new technique is based on targeting objects identified as possible members of the galactic structure under study based on their color-magnitude diagram (CMD) and position in the sky. The objects targeted are those close to the tip of the RGB (TRGB) or brighter. The selection of fainter objects will depend on the telescope and instrument used and the exposure time. These selected candidates are blends of RGB stars, and possibly some asymptotic giant branch (AGB) stars if there is an intermediate-age population in the galactic structure under study \citep[see][for a study of the blending effects on synthetic CMDs]{MD97}. 

We apply this technique to the stellar stream of NGC~4449 \citep{MD12,Rich12}.
Due to the distance to NGC~4449 \citep[$3.82$~Mpc;][]{Annibali08}, the apparent magnitudes of individual RGB stars are too faint to target them spectroscopically and its extremely low surface brightness also prevents us from obtaining integrated light spectroscopy  \citep[the TRGB of NGC~4449's stream is found at $I=24.06$ and its surface brightness is $\mu_g=26.75$~mag~arcsec$^{-2}$;][]{MD12}. However, this stream, and its companion dwarf galaxy NGC~4449, are resolved into individual and blends of stars in deep photometric images and thus it is a perfect target for testing this new technique.

\begin{figure}
\centering
\includegraphics[angle=0,width=9.5cm]{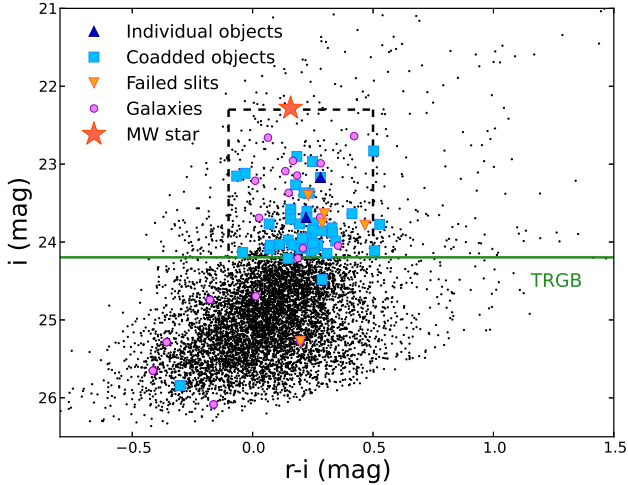}
\caption{Subaru/Suprime-Cam CMD of NGC~4449's stellar stream. The colored symbols indicate the spectroscopic targets for the tidal stream. Dark blue triangles indicate the blends of RGB stars that have enough S/N ($> 2$~pix$^{-1}$) to measure a radial velocity. Light blue squares indicate the blends of RGB stars that are coadded in different spatial bins along the stellar stream to measure a radial velocity. Orange inverted triangles indicate failed slits that do not provide a usable spectrum because of instrumental artifacts. Pink dots indicate background galaxies identified by their emission lines. The red star symbol indicates a Milky Way (MW) star identified by its line-of-sight radial velocity. The green horizontal line indicates the position of the TRGB as estimated by \citet{MD12}. The black dashed lines show our selection box. Bluer and fainter objects than this selection box are added to fill the DEIMOS slitmask as 'filler' targets.}\label{CMD}
\end{figure}

We use our Subaru/Suprime-Cam $r-i$ CMD, which \citet{MD12} used to confirm the distance of the stream, to select candidates in the stellar stream. Our primary targets are objects above the TRGB ($i\leq24.2$) and in the color range $-0.1<r-i<0.5$. A few additional objects with bluer colors and fainter magnitudes are added to fill the DEIMOS slitmask as 'filler' targets, in the sense that they use parts of the slitmask area that would otherwise go unused. Figure \ref{CMD} shows the positions of selected candidates in the CMD. 

\subsection{Observations and data reduction}

\begin{figure*}
\centering
\includegraphics[angle=0,width=7.cm]{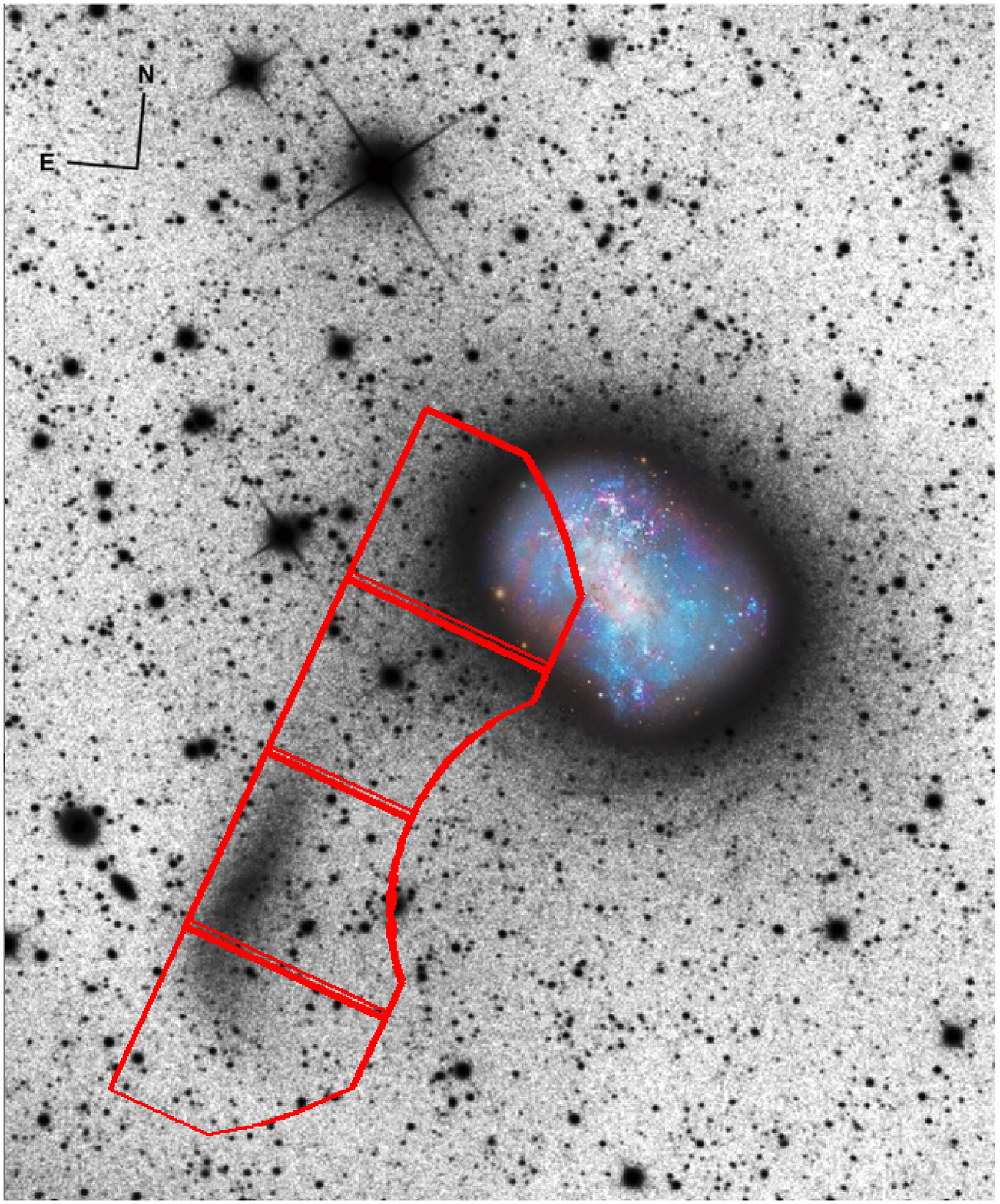}
\includegraphics[angle=0,width=10.8cm]{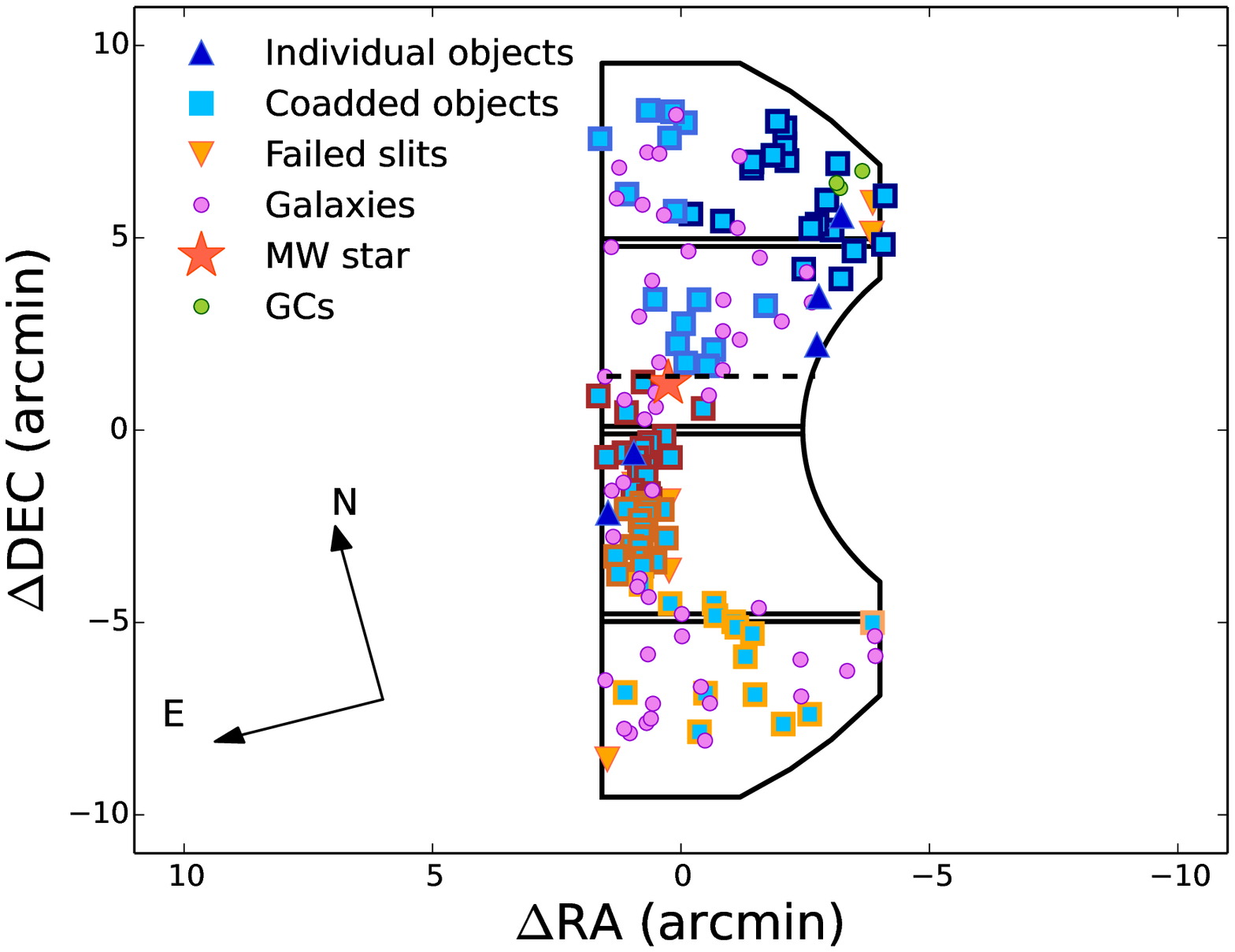}
\caption{Left panel: Keck/DEIMOS footprint used for the spectroscopic observations (in red) overlaid on NGC~4449 and its stellar stream. Perpendicular lines to the longest side of the DEIMOS mask indicate the position of the chip gaps. The image was taken with the BlackBird Remote Observatory and shows a $19.0' \times 24.5'$ ($21\times 27$~kpc) field \citep[adapted from][]{MD12}. Right panel: position of the target objects in the Keck/DEIMOS slitmask. Colors and symbols are the same as in Figure \ref{CMD}. Green dots indicate the position of the three GCs observed in NGC~4449. Blue squares with the same edge color indicate coadded objects. The dashed black line indicates the assumed separation between NGC~4449 and the stellar stream. }\label{deimos_footprint}
\end{figure*}

We used the DEIMOS spectrograph \citep{DEIMOS} located at the Keck~II 10~m telescope in the Mauna Kea Observatory (Hawaii). 
The stream has an elongated morphology that is  $\sim 8$~kpc long by $\sim 1.5$~kpc wide \citep[$\sim 6.4' \times 1.4'$;][]{MD12}. This size is smaller than the Keck/DEIMOS field of view ($16.7' \times 5.0'$). To efficiently fill the slitmask, we added some targets in NGC~4449 itself. These targets were selected in the same way as the targets in the stream. In addition, we selected three globular clusters (GCs) from the catalog by \citet{Strader12} which is based on {\it HST}/ACS photometry. These GCs are brighter than 21.8~mag in the {\it HST}/ACS F555W filter. We did not target more GCs because we gave higher priority to objects selected in the same way as in the stellar stream.
 See Figure \ref{deimos_footprint} for an overlay of the designed  Keck/DEIMOS mask footprint over NGC~4449 and its stream.

The observations were carried out using the 1200~l/mm grating centered at 7800~\AA\ with slit widths of $1.0''$ and the OG550 filter to block shorter wavelength light. All the slits were aligned with the mask position angle (PA$=-12^{\circ}$). This instrumental configuration provides a wavelength coverage of $\sim 6500-9000$~\AA\ with a spectral pixel scale of 0.33~\AA/pixel, and a spectral resolution of 1.4~\AA\ (FWHM) or $R\sim6000$. The observations took place on April 4th, 9th, and 10th 2013. The average seeing conditions were $0.6''$, $0.8''$, and $0.8''$ in FWHM, respectively. The total exposure time for this slitmask was 8400~s split in four individual exposures of 1200~s and 2 of 1800~s.

The raw two-dimensional spectra were reduced and extracted into one-dimensional spectra using the DEIMOS {\sc spec2d} pipeline  designed by the DEEP Galaxy Redshift Survey team \citep{DEIMOSpipeline1,DEIMOSpipeline2} and modified by \citet{SimonGeha07} to optimize the reduction of resolved targets. The main steps in the reduction process consisted of flat-field and fringe corrections, wavelength calibration, sky subtraction, and cosmic ray cleaning. 

The reduced one-dimensional spectrum was obtained by identifying the target in the reduced two-dimensional spectrum and extracting a small window centered on it. The target was identified by finding the peak of the spatial intensity profile obtained by collapsing the two-dimensional spectrum in the wavelength direction. A Gaussian function was fitted to the target and its width was used as the extraction window. The one-dimensional spectrum was obtained by extracting the spectral rows within this window, weighting by the Gaussian function. 

\subsection{Tests to identify the nature of our targets}\label{tests}

The reliability of the results obtained from this technique depends on the nature of the target objects.
Whether our targets mainly consist of several objects with similar luminosity or one luminous and a few fainter objects will affect the accuracy of the radial velocities measured and could dramatically influence the stellar metallicity estimated. If our blends contain objects that are not part of the target galaxy or stream, such as Milky Way (MW) halo and disk stars and background galaxies, the resulting velocities will be affected, too. In the case of MW stars, their similar radial velocity to that of NGC~4449 can broaden the absorption lines and bias the radial velocities and metallicities. In the case of background galaxies only those with absorption lines will affect the resulting measurements. Those with clearly identifiable emission lines within the Keck/DEIMOS spectral window are removed from the sample. The absorption lines of these galaxies will appear at a wavelength that is very different from the absorption lines of NGC~4449 and the stellar stream and thus the radial velocities will not be affected. However, these galaxies could contribute to the continuum, diluting the signal, which results in an underestimation of the metallicity.

Here we perform some tests to analyze the blending effects in our data and to estimate how many MW stars and absorption line background galaxies are expected in our Keck/DEIMOS slitmask.

\subsubsection{Stellar blends}\label{blends}
 
The number of blended stars, as described in Section \ref{description}, depends on the line-of-sight distance, the surface brightness, and the stellar luminosity function. Apart from line-of-sight distance, the other two properties are very different for NGC~4449 versus the stellar stream.  While NGC~4449 is a starburst galaxy with young and intermediate-age stellar populations \citep{Annibali08,Rys11}, the stream, with significantly lower surface brightness, mainly contains old stars ($\sim10$~Gyr) and possibly a handful of intermediate-age AGB stars \citep{MD12}. Thus, the level of blending is expected to be different in NGC~4449 versus the stellar stream. We analyze each of them separately below.

\begin{figure}
\centering
\includegraphics[angle=0,width=9cm]{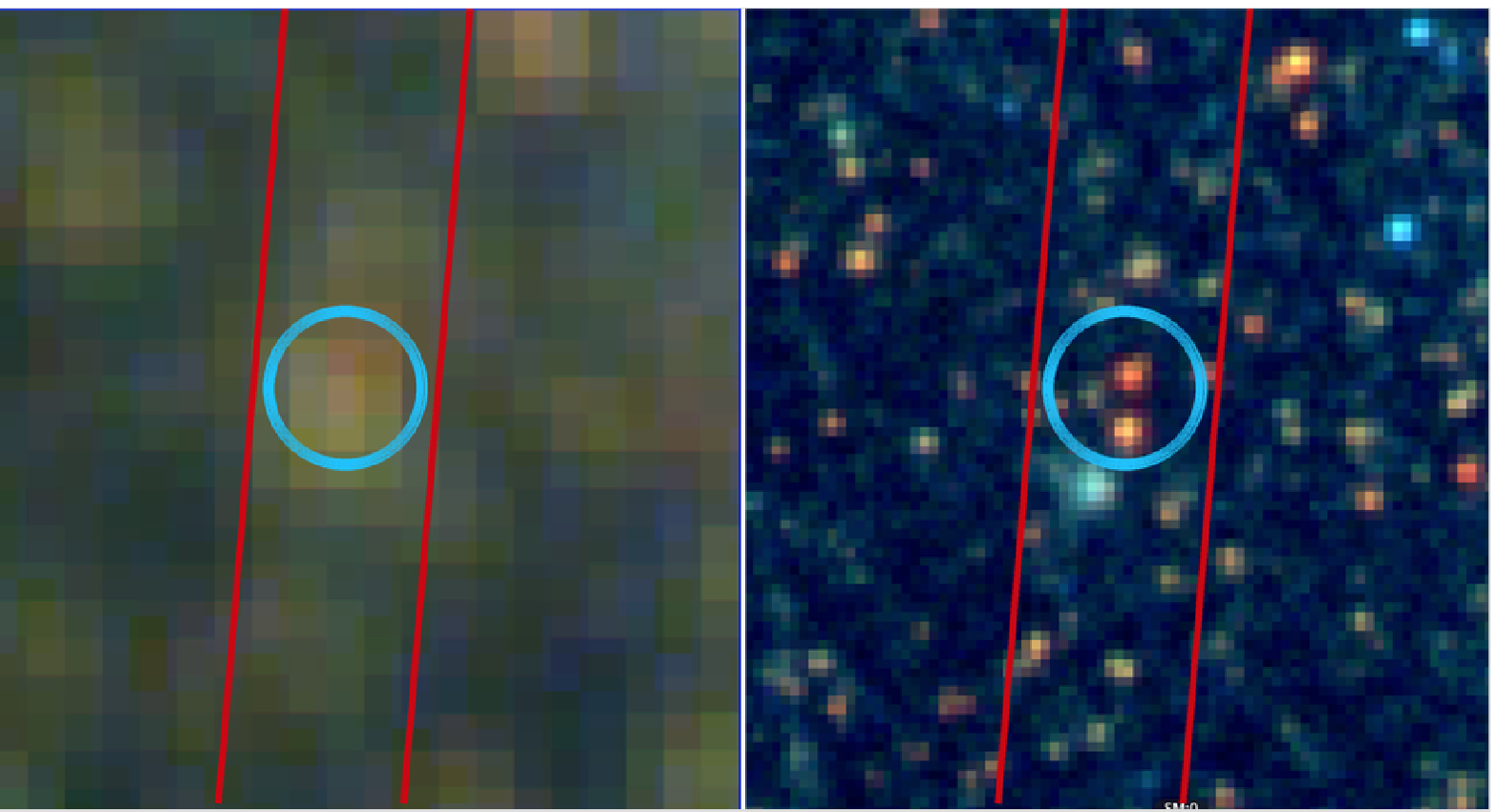}
\includegraphics[angle=0,width=9cm]{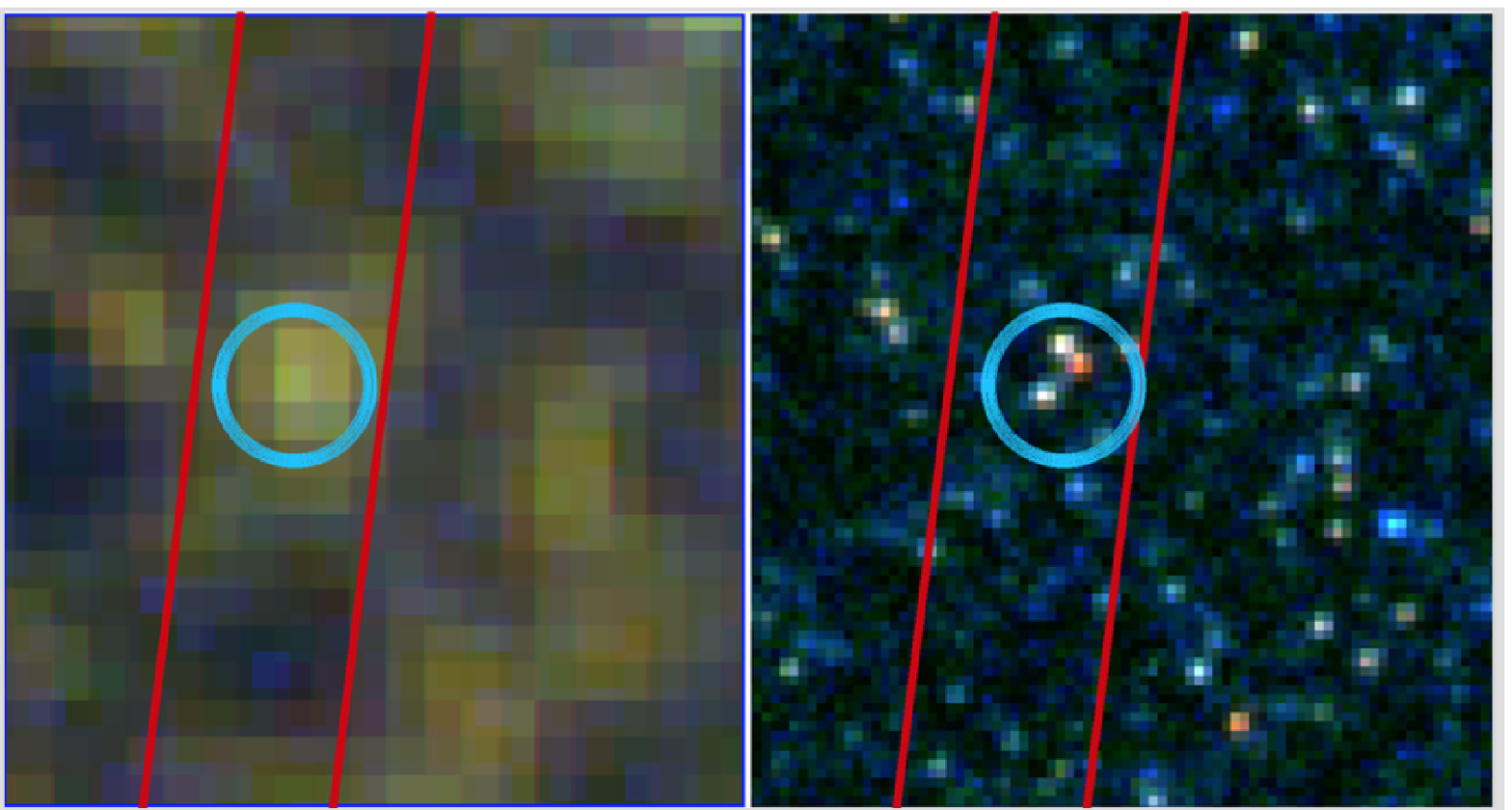}
\includegraphics[angle=0,width=9cm]{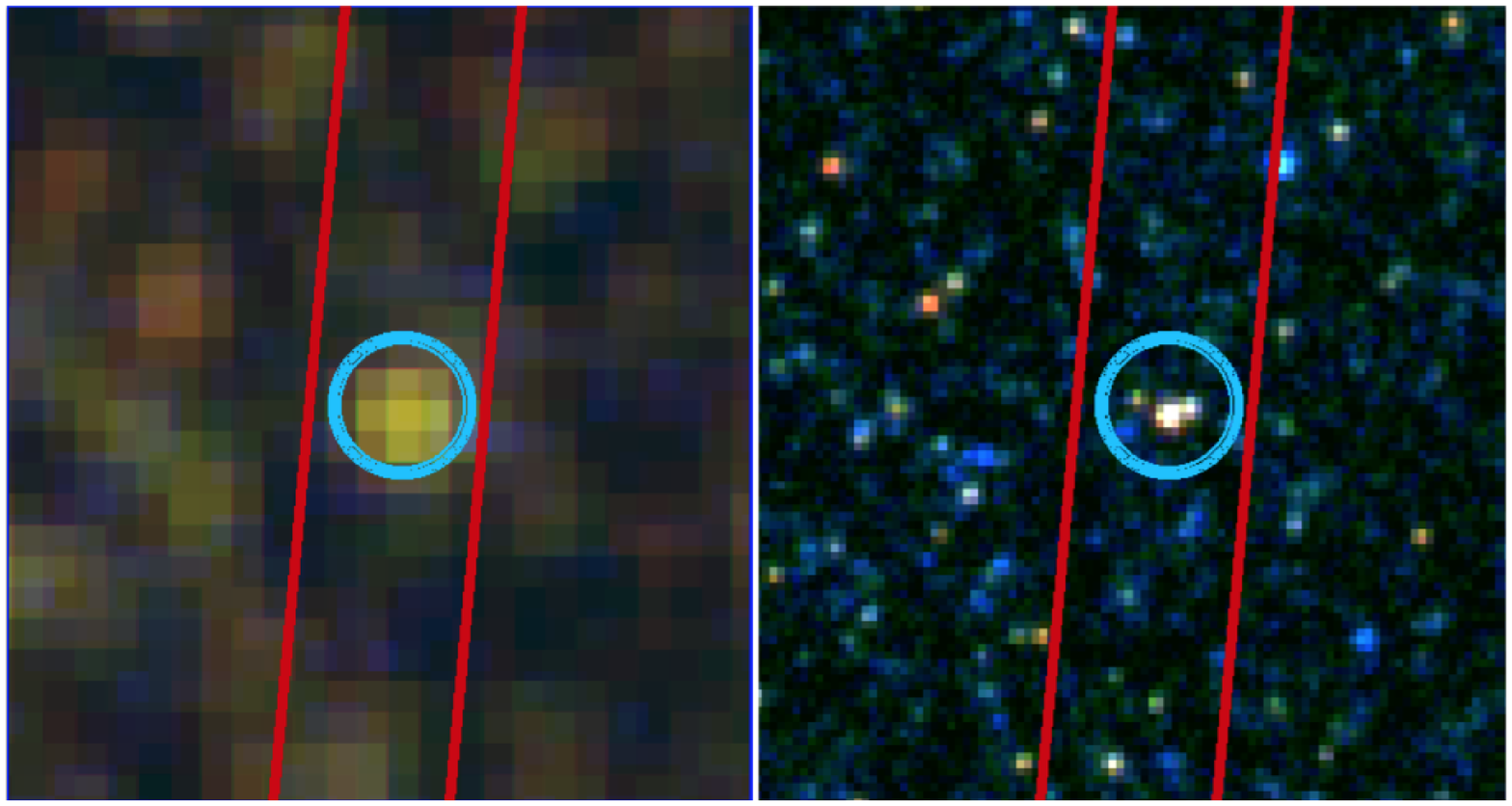}
\caption{Subaru/Suprime-Cam $gri$ color image on the left and {\it HST}/ACS F435W, F555W, and F814W color image on the right for three targets in NGC~4449. The red lines indicate the Keck/DEIMOS slit footprint for the photometric ground based target indicated by the blue circle. The upper panel shows the only slit where our target in the Subaru/Suprime-Cam photometry is a blend of two stars of similar luminosity resolved in the {\it HST}/ACS image. The blue object below our target does not contribute to the light in the wavelength range covered by the spectroscopy ($\sim 6300-9000$~\AA). The remaining of slits are generally composed of one bright star and several fainter stars as shown in the middle and lower panels. Thus, the number of stellar blends that consist of several stars of similar luminosity is expected to be negligible in our sample for NGC~4449. These results are not directly applicable to the stellar stream because of the different luminosity function, see text for more details.}\label{blendsfig}
\end{figure}

We study the blending effects on our NGC~4449 targets by comparing our Subaru/Suprime-Cam photometry with the publicly available {\it HST}/ACS image of the central regions of NGC~4449 \citep{Rys11}.
Only a small region of our Keck/DEIMOS slitmask overlaps with the {\it HST} image of NGC~4449. The {\it HST} image covers only the colored portion of the galaxy image shown in Figure \ref{deimos_footprint}. In that small region, we spectroscopically targeted a dozen objects and three GCs. In the case of the GCs, the surrounding objects are several magnitudes fainter and they do not contribute to the GC spectra.
We examine the blending of the remaining twelve target objects and show an example in Figure \ref{blendsfig}. 

Only in one out of the twelve objects does the stellar blend consist of more than three stars of similar luminosity. This slit is the closest to the center of NGC~4449 in our sample. In this case, the crowding of the stars due to the increase of the stellar density in the central regions of NGC~4449 begins to severely affect the blending. For this reason, we avoided targeting objects in the central regions of the galaxy and targeted objects located more than 1~kpc away from the center of NGC~4449. The S/N of this particular target is too low and we cannot use it to estimate a radial velocity measurement. Therefore, this target is part of the spectral coaddition described in Section \ref{coaddition}. Including or removing this target from our  analysis of radial velocities does not change the results.

In the remaining eleven objects with {\it HST} data, there is only one blend that consists of two stars of similar luminosity. The remaining blends consist of a bright star and a few fainter stars (see Figure \ref{blendsfig}). Thus, the bright star dominates the light in that blend and the radial velocity obtained is not strongly affected by the blending.

There is no {\it HST} image available for the stellar stream, thus, we study its stellar blending by analyzing the expected number of blends in our selection box above the TRGB (see the locus of selected objects in Figure \ref{CMD}). We quantify the expected number of blends by simulating CMDs for streams such as the one near NGC~4449 we have observed here. We do this by taking the photometry of the observed objects in the stream that are below the TRGB and randomly distributing them in space following an exponential profile that spans the total size of the stream ($\sim 1.5\times7$~kpc). We performed 100 such simulations and, for each one, we count the number of RGB stars that appear blended in one seeing disk avoiding duplications. We adopt a FWHM~$0.8''$ as typical seeing for our observations. Then, we calculate the brightness and the color of these seeing disks (the $i$ band magnitude and $r-i$ color). Finally, we count how many of these blends land within our selection box. The median number and standard deviation of these simulations indicate that we should expect $207\pm12$ blends of RGB stars in our selection box.

However, not all the photometric objects are selected to be a spectroscopic target. Using our Subaru/Suprime-Cam catalog of candidate objects in the Keck/DEIMOS field of view ($16.3'\times5'$), we calculate that the fraction of selected objects is $\sim 30\%$. Even though the area of the Keck/DEIMOS slitmask is not filled uniformly and the stellar stream covers a small fraction of the third chip of DEIMOS, counting the chips from North to South in Figure \ref{deimos_footprint}, the selection function is the same because the spatial conflicts do not allow to place more slits in that region. Thus, applying this selection function and taking into account that the stream covers only $\sim 60\%$ of the area of the Keck/DEIMOS slitmask, we expect to have $\sim 37 \pm 2$ blends of RGB stars in our sample of stream objects above the TRGB.

\subsubsection{Background and foreground contamination}\label{contaminants}

We study the kind of objects we target in the stream by estimating the number of contaminants: background galaxies and Milky Way stars. We make this estimation in three independent ways: (1) using our photometric data to estimate the number of objects that land within our selection box; (2) using very deep {\it HST} observations of GOODS-South to estimate the number of compact quiescent galaxies expected in an area the size of the Keck/DEIMOS slitmask. The compact star forming galaxies with emission lines within the wavelength coverage of our instrumental configuration are visually identified and removed from the sample; and (3) using the Besan\c{c}on model to estimate the number of MW stars expected in the line of sight of NGC~4449 in an area the size of the Keck/DEIMOS slitmask.

In the first test, we estimate the number of contaminants by counting the number of objects that land in the selection region above the TRGB in an area of our Subaru/Suprime-Cam image that is not affected by the stream or NGC~4449. Assuming that the distribution of contaminants is uniform, we expect to have 366 contaminants within our selection box. Taking into account the selection function discussed in Section \ref{blends}, we expect to have $\sim 110$ contaminants in our sample and $\sim 66$ of them landing in the area of the stellar stream. These contaminants can be background galaxies, both star forming and quenched, and foreground stars.

In the second test, we use the CANDELS catalog of GOODS-South galaxies by \citet{Guo13} to estimate the number of quenched background galaxies expected in a Keck/DEIMOS slitmask. We select galaxies with a diameter less than or equal to the typical seeing during our  ground-based Subaru/Suprime-Cam observations ($0.5''$). These are identified as point sources in our catalogs. We select quenched galaxies as those with $U-B>1.0$, which roughly represents the red sequence of galaxies \citep[e.g.,][]{Bell04,Faber07}. We finally select objects with $i$ band magnitudes and $r-i$ colors in the same range as our objects (see Figure \ref{CMD}). This selection leads to 0.15 galaxies~arcmin$^{-2}$ in GOODS-South field. Assuming that there is no cosmic variance and taking into account our selection function, we expect to have 3.6 compact quenched galaxies in our sample, of which $\sim 2$ of them can land in the region covered by the stellar stream. 

In the third test, we apply the Besan\c{c}on model \citep{besanconmodel} to the line of sight of NGC~4449, which is located at RA~$=12^{\rm h}28^{\rm m}11.1^{\rm s}$ and DEC~$=+44^{\rm o}5'37''$, or in Galactic coordinates Longitude~$=136.85318^{\rm o}$ and Latitude~$=72.40073^{\rm o}$. We use the $i$ band magnitude range and $r-i$ color range covered by our targets (see Figure \ref{CMD}). The expectation is to have 14 MW stars in the full footprint of Keck/DEIMOS. Applying the same selection function and area coverage to the MW stars, we obtain $\sim 4$ expected MW contaminants. This means that we expect to find $\sim 2$ MW stars in the area where the stellar stream lands. In our radial velocity analysis we find 1 MW star candidate based on its negative radial velocity. The measured radial velocity of NGC~4449 is 207~km~s$^{-1}$ \citep{Sneider92} and the expected velocity dispersion is $\sim 20$~km~s$^{-1}$ based on the Large Magellanic Cloud and other dwarf galaxies of similar luminosity \citep{McConnachie12}. This suggests that an object with a negative velocity is very likely to be a MW star.

\subsubsection{Conclusions from the tests}

NGC~4449 has a much shallower luminosity function than the stellar stream because of its very prominent young population of stars \citep{Rys11}. The young stars in NGC~4449 are brighter than the old stars and stand out in any seeing disk. This makes the blending effects for NGC~4449 negligible. The objects targeted above the TRGB for this galaxy will be dominated by intermediate-age AGB stars.

The stellar stream, on the other hand, has a much steeper luminosity function because it is mainly composed of old stars \citep{MD12}, which means that the bulk of the stars have very similar luminosity. This makes the blending effects in the stellar stream very important. We quantify this blending by counting the number of observed objects in our selection box above the TRGB, which is 651 (based on the CMD shown in Figure \ref{CMD}). Out of these 651 objects, 366 are expected to be background contaminants, 4 are expected to be foreground contaminants, and 207 are expected to be blended RGB stars. Thus, the remaining $\sim 74$ targets could be intermediate-age AGB stars. Taking into account the selection function and the area of the Keck/DEIMOS slitmask covered by the stream, we expect $\sim 13$ intermediate-age AGB stars in our sample of 51 stellar stream objects above the TRGB. The majority of the contaminants are expected to be compact background star-forming galaxies, the number of expected quenched galaxies and MW stars in this area is negligible. These star forming galaxies are easily identified by their emission lines and are removed from our sample.

In summary, our targets in NGC~4449 are mainly individual intermediate-age AGB stars and in the stellar stream our targets are dominated by blends of RGB stars but they can also contain some individual intermediate-age AGB stars.

\subsection{Spectral coaddition}\label{coaddition}

Because of the faintness of the objects targeted, not all the spectra have identifiable absorption lines and the radial velocity measurements are not always reliable. To improve the quality of the data we coadd several spectra in groups. Our grouping scheme is based on the target position on the sky and we require our groups to have a similar number of spectra. As a result, each group has a minimum of fifteen objects and the resulting S/N is at least $\sim 2$~\AA$^{-1}$ in the Calcium triplet region. In the case of the stellar stream, we group objects that are close together along the stream. In the case of NGC~4449's main body, we group objects that are at a similar projected distance with respect to the center of the galaxy (see Figure \ref{deimos_footprint}).

The spectral coaddition process is based on the following steps: (1) rebinning the spectra and their associated uncertainties obtained in the reduction process onto a common wavelength range ($6300-9000$~\AA) -- because the data come from a mask of slits, the spectral range of each slit depends on its position within the mask; (2) renormalizing the fluxes and their associated uncertainties; and (3) adding, pixel by pixel, the fluxes of the normalized rebinned spectra weighted by the scaled version of the rebinned inverse variance (the inverse variance is scaled to match the normalization of the flux). This addition is performed after doing a sigma clipping where those pixels that deviate more than $3\sigma$ from the median are rejected and not added. The distance with respect to NGC~4449 of each coadded spectrum is the average distance of all the objects included in that particular coaddition.
This coaddition technique works best when the internal velocity dispersion of the target galaxy or stream is lower than the velocity uncertainties (typically $20-30$~km~s$^{-1}$; see Table \ref{velgradtable} and Section \ref{vmeasurements}). We assume this to be the case here for NGC~4449 and its stellar stream. The expected velocity dispersion for NGC~4449 is $\sim 20$~km~s$^{-1}$ and for the stream is $<20$~km~s$^{-1}$ given their very low luminosity \citep[e.g.,][]{McConnachie12}.

\begin{figure}
\centering
\includegraphics[angle=0,width=9cm]{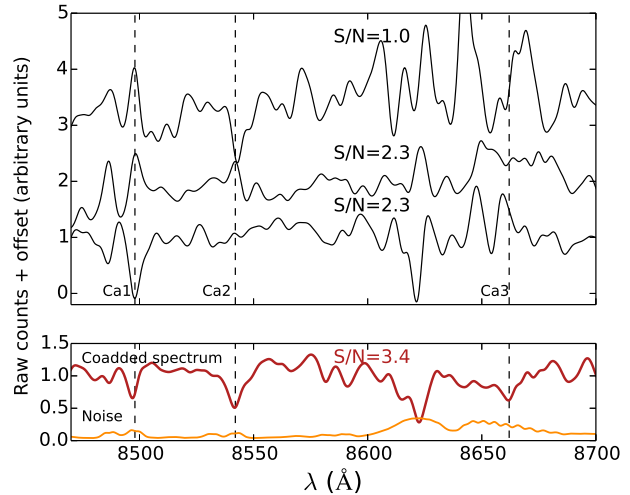}
\caption{Example of the spectral coaddition technique. The upper panel shows the spectra of three of the 19 objects whose coadded spectrum is shown in red in the lower panel. The noise spectrum for this coadded spectrum is shown in orange. The peaks in the noise spectrum correspond to sky line residuals. The spectra have been normalized, smoothed with a Gaussian kernel with $\sigma=2$~pixels weighting by the inverse variance of the spectra, and plotted with a vertical offset. The S/N of each spectrum is measured per \AA. The three Calcium triplet lines are indicated with vertical dashed lines. The red coadded spectrum has higher S/N and shows more spectral features than each individual spectrum. As a result, the radial velocity can be measured in the coadded spectrum while the measurement failed in the individual spectra. }\label{ladderplot}
\end{figure}

Figure \ref{ladderplot} shows an example of a coadded spectrum and some of the individual spectra that go into that coaddition. The individual spectra before being coadded are very noisy and have very few identifiable features. However, when they are coadded together, the features that were very subtle or not identifiable in some of the spectra appear more clearly and can be used to measure a radial velocity.

\section{Radial velocity measurements}\label{vmeasurements}

\begin{figure}
\centering
\includegraphics[angle=0,width=9cm]{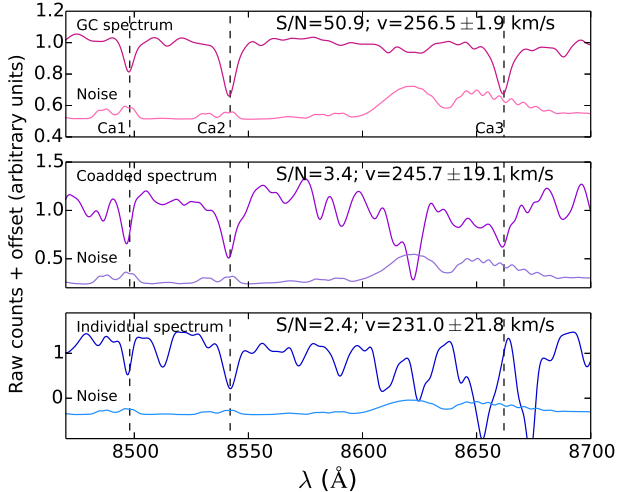}
\caption{Example of rest frame spectrum and noise for a GC (top pannel), coadded spectrum (middle panel), and an individual spectrum (bottom panel). All the spectra are normalized, smoothed with a Gaussian kernel with $\sigma=2$~pixels weighting by the inverse variance of the spectra, and the noise spectrum is plotted with a vertical offset to avoid overlap with the target spectrum. The S/N decreases from top to bottom and, as a consequence, the uncertainties in the measured line-of-sight radial velocities increase.}\label{ladderplot2}
\end{figure}

The line-of-sight radial velocities ($v$) are measured using the penalized pixel-fitting method (pPXF) developed by \citet{PPXF}. This software finds the composite stellar template that best fits the target object. This composite template is created as a linear combination of the stellar templates allowing different weights to minimize the template mismatch problem. 

The stellar templates used are high signal-to-noise (S/N~$>100$) stars observed with Keck/DEIMOS using the same instrumental configuration as the data. To avoid template mismatch, we use 9 stars as templates that include a variety of stellar types, from A to K, luminosity classes, from supergiants to dwarfs, and metallicities, from solar to [Fe/H]$=-3$. 

The uncertainties in the line-of-sight radial velocities are estimated by running 1000 Monte Carlo simulations. In each simulation, the flux of the spectrum is perturbed within a Gaussian function whose width is the flux uncertainty obtained in the reduction process. The radial velocity is measured in each simulation and its uncertainty is defined to be the standard deviation of the Gaussian distribution of all the individual velocities obtained in the simulations. To this random uncertainty we add in quadrature the systematic uncertainty of $1.49$~\kms\ estimated by \citet{Kirby15b} by comparing repeated measurements of the same stars. This systematic uncertainty includes effects that cannot be attributed to the random error, such as uncorrected spectrograph flexure or small errors in the wavelength solution.

The best fit combination of templates is overlaid on the target spectrum and it is visually compared to the distribution of the velocity measurements obtained in the Monte Carlo simulations. The velocity measurements are considered reliable when they are based on spectra where absorption lines are visible after smoothing and when the distribution of the velocity measurements obtained from the Monte Carlo simulations is roughly Gaussian. The adopted velocity is then the center of the roughly Gaussian distribution and the velocity uncertainty is estimated from the width of the distribution. In the spectral range used here, at least two absorption lines are typically identified. The most prominent absorption lines in this spectral range are H$\alpha$ at 6563~\AA\ and the Ca lines at 8498, 8542, and  8662~\AA. However, the radial velocities of NGC~4449 and the stream make the Ca1 and Ca2 lines coincide with two prominent sky lines, which affects our radial velocity measurements. Figure \ref{ladderplot2} shows examples of spectra for which we retrieve a reliable radial velocity measurement as a function of S/N. This figure shows a spectrum of a globular cluster, a coadded spectrum, and an individual spectrum that could be an individual intermediate-age AGB star or a blend of RGB stars.

Those objects for which a radial velocity cannot be retrieved or is considered unreliable following the criterion above are coadded following the steps described in Section \ref{coaddition}. These objects typically have spectra with S/N~$< 2$~\AA$^{-1}$. At least fifteen objects need to be coadded to obtain enough S/N to measure a reliable radial velocity. Table \ref{velgradtable} shows the radial velocities obtained for all the objects in our sample.

\begin{table*} 
\begin{center} 
\caption{Properties of the Spectroscopic Targets \label{velgradtable}} 
\resizebox{15cm}{!}{ 
\begin{tabular}{c|c|c|c|c|c|c|c} 
\hline \hline 
RA         &    DEC    &    $D$   &   $r$   &   $i$   &  F555W   &  $v$         & Object    \\ 
(hh:mm:ss) & (dd:mm:ss)&   (kpc)  &  (mag)  &  (mag)  &  (mag)   &  km~s$^{-1}$ &           \\ 
   (1)     &  (2)      &   (3)    &   (4)   &   (5)    &   (6)   &   (7)        &    (8)    \\ 
\hline 
\multicolumn{8}{c}{NGC~4449} \\ 
\hline 
12:28:25.84 & 44:05:54.04 &  3.00 & 22.88 & 22.51 & --- & 249.1$\pm$ 17.5 & ind\\ 
12:28:27.66 & 44:03:48.24 &  3.90 & 24.32 & 23.56 & --- & 254.3$\pm$ 20.5 & ind\\ 
12:28:27.81 & 44:02:32.84 &  4.80 & 24.12 & 23.47 & --- & 223.7$\pm$ 27.3 & ind\\ 
--- & --- &  2.70 & [25.01,23.00] & [24.40,22.61] & --- & 222.7$\pm$ 16.6 & coadd (19)\\ 
--- & --- &  5.30 & [25.25,23.11] & [24.46,22.60] & --- & 167.8$\pm$ 66.7 & coadd (15)\\ 
12:28:16.55 & 44:05:36.50 &  1.10 & --- & --- & 19.64 & 254.7$\pm$  2.2 & GC (A7)\\ 
12:28:18.79 & 44:05:19.50 &  1.60 & --- & --- & 19.66 & 124.0$\pm$  2.6 & GC (A6)\\ 
12:28:18.90 & 44:05:28.30 &  1.60 & --- & --- & 21.84 & 256.8$\pm$ 10.5 & GC (A31)\\ 
12:28:38.24 & 44:01:33.96 &  7.00 & 22.44 & 22.28 & --- & $-$103.0$\pm$ 48.2 & MW\\ 
\hline 
\multicolumn{8}{c}{Stellar Stream} \\ 
\hline 
12:28:42.57 & 43:59:44.48 &  9.10 & 23.45 & 23.17 & --- & 258.0$\pm$ 21.7 & ind\\ 
12:28:44.64 & 43:58:11.03 & 10.60 & 23.91 & 23.69 & --- & 239.7$\pm$ 17.7 & ind\\ 
--- & --- &  8.80 & [24.62,23.09] & [24.15,22.83] & --- & 216.8$\pm$ 58.5 & coadd (18)\\ 
--- & --- & 10.90 & [24.33,23.09] & [24.08,22.97] & --- & 215.8$\pm$ 39.5 & coadd (16)\\ 
--- & --- & 13.90 & [25.92,21.23] & [26.09,20.26] & --- & 223.1$\pm$ 42.9 & coadd (15)\\ 
\hline 
\end{tabular} 
} 
\end{center} 
\tablecomments{Columns 1 and 2: Coordinates in J2000. Column 3:
  Projected distance to the center of NGC~4449. Columns 4 and 5:
  Apparent $r$ and $i$ band magnitudes from the Subaru/SuprimeCam
  photometry. For the coadded objects, we indicate the apparent
  magnitude range of the sources that are coadded. Column 6: Apparent
  F555W magnitudes from the {\it HST}/ACS catalog by
  \citet{Strader12}. These magnitudes are only available for the
  globular clusters. Column 7: Heliocentric radial velocity. Column 8:
  Object type: {\it Ind} indicates individual object; {\it coadd}
  indicates coadded objects, where the number of coadded objects is
  indicated within parentheses; {\it GC} indicates globular clusters,
  where the
  name within parentheses corresponds to the name in the catalog by \citet{Strader12}; and {\it MW} indicates that the object is likely to be a Milky Way star based on its radial velocity.} 
\end{table*}


Three globular clusters in NGC~4449 are within our sample of targets. We estimate their radial velocities following the same method described above. The radial velocity uncertainties for these three GCs are significantly smaller than for any other object in our sample because of their higher S/N due to being the brightest objects in the sample. The heliocentric radial velocity that we measure for the GC A6 does not agree within the $1\sigma$ uncertainty with the value measured by \citet[][$v=103.5\pm1.5$]{Strader12}. In the case of \citet{Strader12}, the value obtained is even farther away from the systemic velocity of NGC~4449 than ours.

\begin{figure}
\centering
\includegraphics[angle=0,width=9cm]{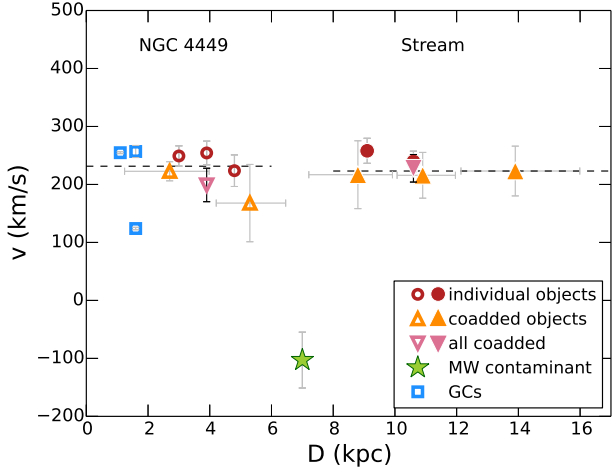}
\caption{Velocity gradient in the main body of NGC~4449 and along the stream. The open symbols indicate objects likely belonging to NGC~4449 and the filled symbols indicate objects likely belonging to the stream. The center of the stream is found at $\sim 11$~kpc from NGC~4449's center and its apparent turning point at $\sim 14$~kpc. The horizontal error bars indicate the range of distances of the individual objects included in each coadded spectra. The dashed lines show the median velocity for NGC~4449 and the stream obtained independently. The pink inverted triangles show the radial velocity measured in the spectrum obtained after coadding all the objects in NGC~4449 and the stream, respectively, without the GCs. The median velocity and these global measurements are in good agreement. NGC~4449 and the stream have similar radial velocities to within measurement uncertainties, consistent with what may be expected if the tidal stream is gravitationally bound to NGC~4449.}\label{vgrad}
\end{figure}

Figure \ref{vgrad} shows the resulting radial velocities as a function of position with respect to the center of NGC~4449. The radial velocities for the GCs and the Milky Way star identified by its radial velocity are also plotted in this diagram. Although the stream is superposed on the main body of NGC~4449, the probability of targeting objects that belong to the stream decreases as we approach NGC~4449's central regions. For this reason, we assume that any object targeted in the region where the surface brightness of NGC~4449 is higher than the surface brightness of the stream belongs to NGC~4449's main body. This roughly corresponds to the objects targeted in the upper 1.5 chips of the Keck/DEIMOS footprint shown in Figure \ref{deimos_footprint}, the inner $\sim 7$~kpc of the galaxy. The apparent gap at distances between 6 to 8~kpc from NGC~4449 corresponds to the region of very low density of stream candidates. However, the gap is not real, but is a consequence of the coaddition procedure (see Figure \ref{deimos_footprint}). All the objects in the region $6-8$~kpc are coadded into the spectra that lead to the data points in Figure \ref{vgrad} located at 5.5~kpc and 8.3~kpc. These distances are calculated as the mean distance of all the target objects included in each coadded spectrum.

The heliocentric line-of-sight radial velocity of NGC~4449 is estimated as the median of all the objects in the inner $\sim 7$~kpc ( $227.3\pm10.7$; see Tables \ref{velgradtable} and \ref{properties}). Its value is consistent with the value obtained averaging the neutral hydrogen velocity measured in the high and low velocity sides of the profile \citep[$207$~\kms;][]{Sneider92}. In addition, the radial velocities obtained for NGC~4449 along its inner 7~kpc are consistent with the values obtained for the HII regions that are in the same area of the galaxy \citep{Hartmann86,Hunter02}.  We estimate the heliocentric line-of-sight radial velocity for the stream as the median of all objects with distances from NGC~4449 larger than $\sim 8$~kpc. We find a value of $225.8\pm16.0$~\kms\ (see Table \ref{properties}). See Section \ref{velgrad} for a discussion about the velocity gradients.


\begin{table}
\begin{center}
\caption{Velocity and Metallicity Properties of NGC~4449 and its Stream\label{properties}}
\begin{tabular}{c|c|c|c}
\hline \hline
   Object      &     $v$       & $\Sigma$Ca    &  [Fe/H]\\
                   &    (\kms)    &  (\AA)             &    (dex) \\
     (1)          &        (2)      &      (3)         &   (4)           \\     
\hline
NGC~4449-East   &   $227.3\pm10.7$  & $8.7\pm0.9$    & ---\\
Stream         &    $225.8\pm16.0$  & $4.9\pm0.7$   & $-1.37\pm0.41$\\
\hline
\end{tabular}
\end{center}
\tablecomments{Column 1: object. Column 2: heliocentric corrected line-of-sight radial velocity calculated as the median of all the velocities obtained for the object indicated in column 1. Column 3: total Calcium triplet equivalent width. Column 4: metallicity obtained by coadding individual and blends of candidate RGB stars. }
\end{table}

\section{Metallicity measurements}\label{Zmeasurements}

The stellar metallicity [Fe/H] can be estimated by using the near-infrared Calcium triplet (CaT) absorption lines. The technique of spectral coaddition has successfully been used for individual low S/N RGB stars \citep[e.g.,][]{Collins13,Yang13}. However, the technique used here combines this traditional spectral coaddition with a different kind of coaddition, the coaddition by nature, we target bright objects which are potentially blends of stars. Thus, to estimate the metallicity of the stream, we assume that the stars involved in the blending are RGB stars and that the number of intermediate-age AGB stars is negligible as estimated in Section \ref{tests}.

We measure the equivalent widths (EWs) of the CaT lines and use the \citet{Carrera13} metallicity calibration to transform the observed CaT EWs and $I$ band luminosities of the objects involved into [Fe/H]. The EW is usually estimated by fitting the target absorption line with a function and integrating over the area of the resulting fit. The shapes of the CaT lines are well described by a Gaussian function for low metallicity systems while they are better described by a combination of a Gaussian and a Lorentzian function for high metallicity systems \citep{Cole04,Carrera07,Saviane12}. Although using a combination of Gaussian and Lorentzian functions is the most accurate procedure for all metallicities, this methodology is not stable when working with low S/N spectra because the imperfections of the sky subtraction can modify the shape of the 8662~\AA\ line \citep[S/N~$<25$~pixel$^{-1}$;][]{Ho15}. In the case of low S/N spectra it is better to fit the CaT lines with a Gaussian function and then use a correction that transforms it into the value that would have been obtained if a Gaussian and a Lorentzian function were used. The correction we use is:

\begin{equation}\label{correction}
{\rm EW}_{G+L} = 1.114(\pm0.01)\times {\rm EW}_G
\end{equation}

\noindent where EW$_G$ is the EW measured by fitting a Gaussian function and EW$_{G+L}$ is the EW obtained by fitting a Gaussian plus a Lorentzian functions. This correction, based on the EW definition for the CaT lines by \citet{Armandroff91}, was obtained by measuring the EW$_G$ and the EW$_{G+L}$ in several hundred high S/N (S/N~$>25$~pixel$^{-1}$) RGB stars in globular clusters \citep{Ho15}.

We measure the EW of the CaT lines for those objects in NGC~4449's main body and in the stellar stream independently. Following the spatial distribution of objects shown in Figures \ref{deimos_footprint} and \ref{vgrad}, NGC~4449's main body consists of the coaddition of all the spectra in the inner 7~kpc, and the stream consists of the coaddition of all the spectra at larger distances from the center of the galaxy. The spectral coaddition technique followed is described in Section \ref{coaddition}. There are a total of 37 spectra coadded in the galaxy's main body and 51 in the stream. The resulting spectra have a S/N of 5.2~\AA$^{-1}$ and 6.6~\AA$^{-1}$, respectively, and a mean $I$ band magnitude of 23.1 and 23.8, respectively. The $i$ band from our photometry is transformed into the $I$ band following the relation of \citet{Jordi06} and assuming a median $r-i$ color of 0.55 based on the CMD shown in Figure \ref{CMD}.

\begin{figure}
\centering
\includegraphics[angle=0,width=9cm]{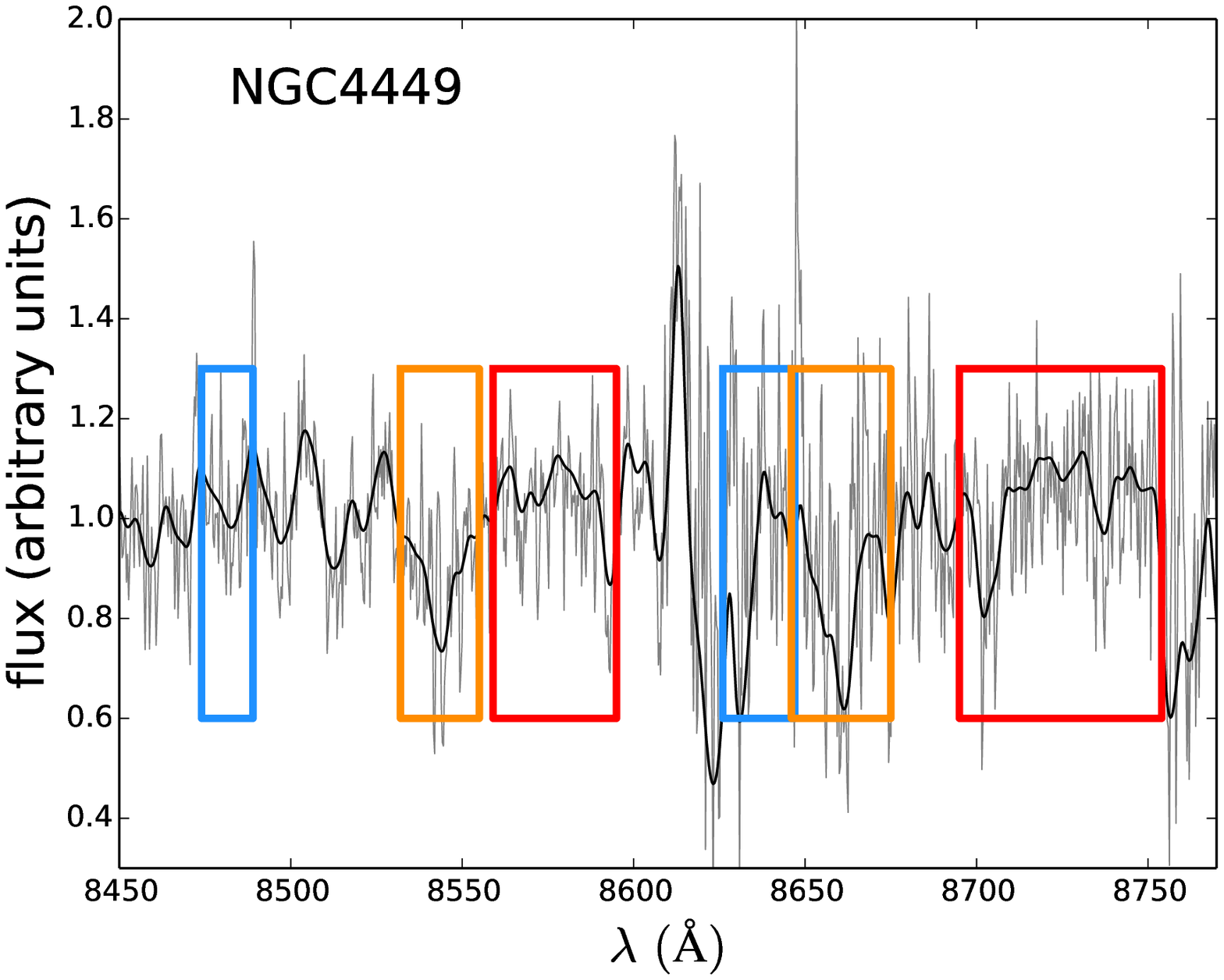}
\includegraphics[angle=0,width=9cm]{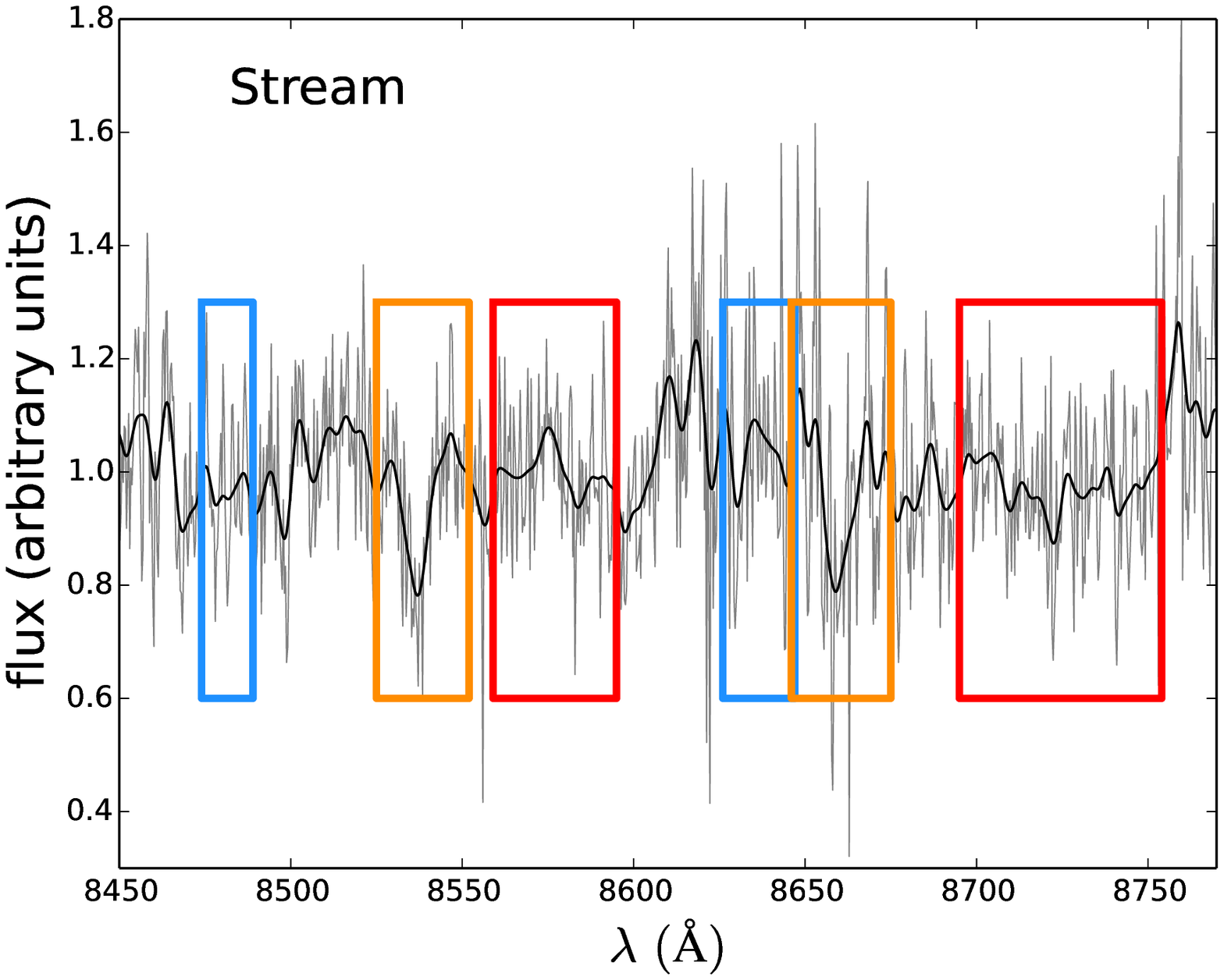}
\caption{Spectra showing the Calcium lines. The resulting coadded spectrum is shown in grey. The coadded spectrum smoothed by a Gaussian kernel of 1.5~pixel weighting by the inverse variance of the spectrum is shown in black. The regions used to fit a Gaussian function to the two more prominent Ca lines are shown in orange. The continuum regions used to obtain the best fit Gaussian function are indicated by a blue and red boxes at both sides of the fitted line. The bandpass and continuum regions used to fit the Ca absorption line are those defined by \citet{Armandroff91} and are simultaneously constrained in the fit. Upper panel: spectrum based on the coaddition of 37 individual spectra that belong to NGC~4449. The S/N of this spectrum is 5.2~\AA$^{-1}$. Lower panel: spectrum based on the coaddition of 51 spectra of blends of RGB stars that belong to the stream. The S/N of this spectrum is 6.6~\AA$^{-1}$.}\label{Gaussfit}
\end{figure}

We smooth our coadded spectra using a Gaussian kernel of 1.5~\AA\ to match the resolution of the line and continuum regions defined by \citet{Armandroff91} for the Ca lines at 8542~\AA\ and 8662~\AA. Following the method of \citet{Ho15}, we fit a Gaussian function to each line independently and apply the correction indicated in Equation \ref{correction}. Figure \ref{Gaussfit} shows the spectral regions used to make the fit. We then determine the total EW of the Ca lines as the unweighted sum of the two lines:

\begin{equation}\label{sigCa}
\Sigma {\rm Ca} = {\rm EW}_{8542} + {\rm EW}_{8662}
\end{equation}

We estimate the uncertainty of $\Sigma$Ca by running 1000 Monte Carlo simulations where every pixel in the spectrum is randomly perturbed within a Gaussian function whose width is equal to the flux uncertainty in that pixel obtained during the data reduction process. In each simulation, we calculate the new EW$_{8542}$ and EW$_{8662}$, and the uncertainty in these two parameters is the standard deviation of all the simulations. The uncertainty in $\Sigma$Ca is then the square root of the quadratic sum of the uncertainties in EW$_{8542}$ and EW$_{8662}$ (see Table \ref{properties}).

The conversion from $\Sigma$Ca to [Fe/H] depends on the luminosity of the individual star. The relation between $\Sigma$Ca  and [Fe/H] is linear when the luminosity indicator used is the magnitude of the star above the horizontal branch (HB). This is because the relation between $\Sigma$Ca and the magnitude above the HB is nearly linear for tracks of equal metallicity. However, the HB is difficult to determine for objects at large distances, we use instead the absolute $I$ band magnitude of the RGB star, which is also a robust parameter to estimate [Fe/H] \citep{Carrera07}. To transform $\Sigma$Ca and the magnitude of the RGB star into [Fe/H] we use the conversion by \citet{Ho15}, which is specifically obtained for a $\Sigma$Ca value based on the two Ca lines at 8542~\AA\ and 8662~\AA\ and has been tested for the same procedure described here:

\begin{eqnarray}\label{metallicity}
{\rm [Fe/H]} &=& -3.51+0.12\times M_I+0.57\times \Sigma{\rm Ca} \nonumber \\
                   & -& 0.17\times \Sigma{\rm Ca}^{-1.5} +0.02\times \Sigma{\rm Ca}\times M_I
\end{eqnarray}

In our case, $\Sigma$Ca is measured in the spectra that results from coadding all the targets in NGC~4449 and the stream independently (see Figure \ref{Gaussfit}). These targets are dominated by intermediate-age AGB stars in NGC~4449 and blends of RGB stars in the stellar stream. Thus, while for NGC~4449 we use the traditional coaddition technique, for the stellar stream we use this traditional technique to coadd the upward fluctuations of the SBF, thus, we coadd blends of RGB stars. 

We cannot estimate the metallicity of NGC~4449 because the calibration in Equation \ref{metallicity} is specific to RGB stars. In the case of the stellar stream, the absolute magnitude $M_I$ in Equation \ref{metallicity} refers to the average $I$ band magnitude of the objects included in the resulting spectrum. The simulations described in Section \ref{blends} show that the typical blending consists of two RGB stars with a luminosity ratio in the $I$ band of $0.64\pm0.20$. We use this ratio to estimate the average absolute magnitude $M_I$.

The uncertainty in the metallicity is obtained by propagating the errors in $\Sigma$Ca and $M_I$, where the uncertainty in $M_I$ is assumed to be the square root of the quadratic sum of three components: 1) the square root of the quadratic sum of the random and systematic $I$ band uncertainties estimated by \citet[][]{MD12}; 2) the uncertainty of the typical luminosity ratio of the RGB stars involved in the blending from the simulations; and 3) the uncertainty in the distance to the stream estimated by \citet[][]{MD12}. However, due to the low S/N of our spectra, the [Fe/H] uncertainty is dominated by the uncertainty in $\Sigma$Ca.

The stellar metallicity estimated for the stellar stream is in agreement with the value obtained from overplotting isochrones over its CMD, which provides a metallicity range of $-0.68>$[Fe/H]$>-1.27$, that depends on the age of the isochrones considered \citep{MD12}. The metallicity obtained with our spectroscopic coaddition method is [Fe/H]~$=-1.37\pm0.41$ (see Table \ref{properties}).

\begin{figure}
\centering
\includegraphics[angle=0,width=9cm]{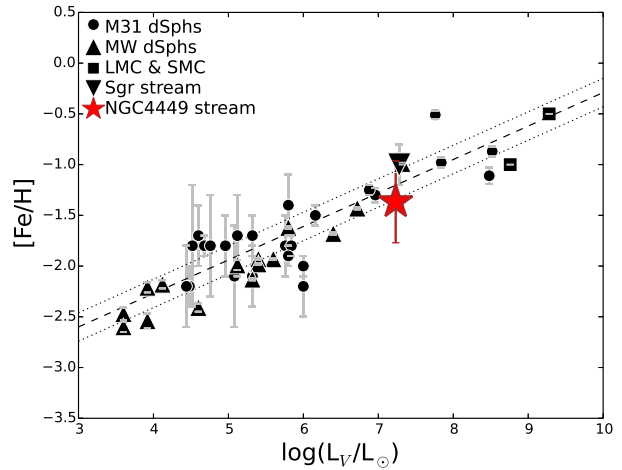}
\caption{Metallicity-luminosity relation for dwarf galaxies. The red star indicates the metallicity measured using the coaddition of blends of RGB stars described in this work for NGC~4449's stellar stream. For comparison, we show the metallicity measurements for the Local Group dwarf galaxies available in the literature. The two black squares indicate the Large and Small Magellanic Clouds \citep[LMC, SMC;][]{Carrera08a,Carrera08b}. The black triangles indicate the Milky Way dSphs \citep{Kirby11}. The black dots indicate the M31 dSphs, dEs, and M32 \citep{Collins13,Ho15}. The inverted black triangle indicates the Sagittarius stream \citep{Mateo98LG}}\label{metallumin}
\end{figure}

\section{Discussion}\label{discussion}

\subsection{Velocity gradient along NGC~4449 and the stellar stream}\label{velgrad}

We measure the line-of-sight radial velocity in the inner $\sim 7$~kpc of NGC~4449 and along $\sim 8$~kpc of the stellar stream. The heliocentric median line-of-sight radial velocities of NGC~4449 ($227.3\pm10.7$~km~s$^{-1}$) and the stream ($225.8\pm16.0$~km~s$^{-1}$) are very similar, which suggests that they are likely gravitationally bound. 
We find a flat stellar velocity gradient on the East side of NGC~4449 and along the stellar stream, however, the uncertainties in our velocity measurements are large. If the orbit of the stream is close to the plane of the sky, a flat velocity gradient along the stellar stream is expected. However, if the orbit of the stream has an inclination with respect to the plane of the sky and, in the most extreme case, the stream has a radial trajectory (it is radially falling into the galaxy), then we can estimate the expected velocity gradient following the calculations of \citet[][$\Delta v/v_c=\Delta r/r$]{MerrifieldKuijken98}. The circular velocity ($v_c$) at the apocentric radius ($r$~14~kpc, the apparent turning point of the stream) is 62~km~s$^{-1}$ \citep{MD12}, and the region along which we want to calculate the velocity gradient is $\Delta r=7$~kpc. These numbers give an expected velocity gradient of $\Delta v=35$~km~s$^{-1}$. Our velocity uncertainties are too large to be able to measure that velocity gradient.

This kinematic information is key for developing detailed models of the collision of the progenitor of the stream and NGC~4449, thus, these uncertainties need to be improved to better constrain the possible lack of velocity gradient in the stellar stream. That can be achieved either by exposing longer and obtaining higher S/N spectra or by designing multiple slitmasks to obtain spectra for more objects in NGC~4449 and the stream.

\subsection{Metallicity of NGC~4449's stellar stream}

The metal content of galaxies follow a sequence with their stellar mass and, thus, their luminosity \citep[e.g][]{Gallazzi05}. More massive galaxies are more metal rich than less massive galaxies. This sequence extends to the dwarf galaxy regime \citep{Kirby08b,Kirby11} and seems to be universal for quenched and star forming galaxies \citep{Gallazzi05,Kirby13}. In the case of streams, this metallicity-luminosity relation can be used to constrain the total luminosity of the progenitor galaxy. Systems that are tidally stripped would move horizontally to the left in the metallicity-luminosity plane because their luminosity would gradually decrease as stars are stripped, but the metallicity would remain unaffected because there is no new star formation.

Figure \ref{metallumin} shows the metallicity-luminosity relation for dwarf galaxies. We use as reference the average metallicity based on individual RGB stars for Milky Way dSphs, M31 faint dSphs, and M31 bright dSphs, dEs, and M32 \citep[][respectively]{Kirby11,Collins13,Ho15}. We also include the average stellar metallicity values for the Large and Small Magellanic Clouds \citep[LMC, SMG;][]{Carrera08a,Carrera08b} and the Sagittarius stream \citep{Mateo98LG}. The metallicity we obtain by using this new technique for the stellar stream follows the metallicity-luminosity relation of dwarf galaxies in the Local Group. The stream's stellar metallicity is consistent with the metallicity of the Fornax dwarf spheroidal galaxy and the Sagittarius stream and both of them have a luminosity that is very similar to the luminosity of NGC~4449's stellar stream. Due to the large metallicity uncertainty, it is difficult to constrain whether the stripping of stars experienced by the stream has been moderate or severe, and whether the accretion event has been recent or happened a long time ago.

\section{Summary}\label{summary}

We present a new spectroscopic technique to measure the stellar kinematics and metallicities of diffuse objects that are beyond the Local Group. 
This technique consists of targeting objects that are above the tip of the RGB. These objects are a mixture of some AGB stars, if an intermediate-age population is present in the target galaxy or stream, and blends of RGB stars that appear brighter because of their spatial superposition in the sky. Given the brightness of these objects ($I \gtrsim 23.5$~mag) we can measure their line-of-sight radial velocities and, coadding their individual spectra, we can also estimate the metallicity of the target galaxy or stream. The metallicity measurements are limited to old stellar populations where intermediate-age AGB stars are not present or their number is negligible. This is because of the lack of a calibration to transform CaT EWs into metallicities. This limitation will no longer exist when these calibrations are established.

We apply this technique to the nearby star forming dwarf galaxy NGC~4449 and its stellar stream located at 3.8~Mpc \citep[][]{Annibali08}. We measure the line-of-sight radial velocity along the galaxy and the stream and find flat radial velocity gradients for both systems, within our velocity uncertainties and the fact that for the galaxy we only sample its East side and it is not along the semimajor axis. We also find that the median heliocentric velocity of both systems are very similar to each other, $V=227.3\pm10.7$~\kms\ for NGC~4449 and $V=225.8\pm16.0$~\kms\ for the stream, which suggests that the two systems are gravitationally bound. We estimate the global metallicity of the stellar stream because it is mainly composed old stars, while our targets for NGC~4449 seem to be mainly intermediate-age AGB stars. Our measured metallicity for the stream is [Fe/H]~$=-1.37\pm0.41$, which is consistent with the metallicity-luminosity relation of Local Group dwarf galaxies. However, the uncertainty of the metallicity is too large to constrain whether the progenitor of the stream was a dwarf galaxy with properties very similar to those dwarfs found in the halo of the Milky Way or M31, or, on the contrary, it was a larger galaxy that was severely stripped by NGC~4449, as it happened for the Sagittarius stream which has the same luminosity and metallicity as the stream.

We demonstrate that this new SBF spectroscopic technique is a powerful method for studying the kinematics and metallicities of the wealth of dwarf faint satellites and streams that are currently being discovered in the local universe. We will apply this same technique to the newly discovered dwarf satellites of M81 located at $\sim 3.8$~Mpc  \citep{Chiboucas09,Chiboucas13} and to the dwarf galaxies and streams that we are finding  within the PISCeS survey \citep[Panoramic Imaging Survey of Centaurus and Sculptor;][]{Sand14,Crnojevic14,Crnojevic15,etj16}. These data provide ingredients for detail modeling of the collision between the progenitor galaxy and the host galaxy, and also provide observational constraints on the host's potential well and on the properties and orbit of the progenitor of the streams and satellites.

\acknowledgments

E.T. thanks Josh Simon, Evan Kirby, Denija Crnojevi$\acute{{\rm c}}$, Guillermo Barro, and David Sand for very helpful discussions. The authors thank the referee for the useful comments that helped improve this manuscript. E.T. and P.G. acknowledge the NSF grants AST-1010039 and AST-1412504. J.B. acknowledges support from the NSF grants AST-1211915 and AST-1109878.
The spectroscopic data presented herein were obtained at the W.M. Keck Observatory, which is operated as a scientific partnership among the California Institute of Technology, the University of California and the National Aeronautics and Space Administration. The Observatory was made possible by the generous financial support of the W.M. Keck Foundation. The photometric data presented herein are based on observations obtained with the Subaru Telescope, operated by the National Astronomical Observatory of Japan, via Gemini Observatory time exchange (GN-2010B-204). The authors wish to recognize and acknowledge the very significant cultural role and reverence that the summit of Mauna Kea has always had within the indigenous Hawaiian community.  We are most fortunate to have the opportunity to conduct observations from this mountain.

\bibliographystyle{aa}
\bibliography{references}{}


\end{document}